\documentclass[final,5p,times,compress,twocolumn]{elsarticle}
\usepackage{amssymb}
\usepackage{amsmath}
\usepackage{lipsum}
\usepackage{xcolor}
\usepackage[colorlinks=true, allcolors=blue]{hyperref}
\usepackage{orcidlink}
\usepackage[normalem]{ulem}
\usepackage{enumitem}
\usepackage{subcaption}
\usepackage{lineno}

\journal{Physics Letters B}

\begin{document}

\begin{frontmatter}

\title{$K^-$--Driven Direct Urca Cooling in Rotating Neutron Stars: A Bayesian Study}

\author{Aprajita Shrivastava\corref{cor1}}
\author{Debanjan Guha Roy}
\author{Sarmistha Banik}
\cortext[cor1]{p20230071@hyderabad.bits-pilani.ac.in}
\affiliation{organization={Department of Physics, BITS Pilani, Hyderabad Campus},%Department and Organization
            % addressline={}, 
            city={Hyderabad},
            postcode={500078}, 
            state={Telengana},
            country={India}}

\begin{abstract}
We investigate the onset of antikaon ($K^-$) condensation and its implications for the equation of state (EoS) and  cooling of neutron stars (NSs) within density-dependent relativistic mean-field parametrisations DD2 and MPE. Treating the antikaon - nucleon optical potential ($U_K$) as a free parameter in the range $[-180,-60]$ MeV, we constrain it using Bayesian inference with NICER mass-radius observations of PSR J0030+0451 and PSR J0740+6620. The inferred posterior distributions favour strongly attractive in-medium $K^-$ interactions, while their broad widths indicate only weak constraints on $U_K$ by astrophysical observations.
More attractive values of $U_K$ lead to an earlier onset of $K^-$ condensation, enhanced softening of the EoS, and lower Direct Urca (DU) threshold densities. The condensation threshold is systematically lower in DD2 than in MPE, while finite entropy further promotes the onset of rapid cooling. The $K^-$--induced enhancement of the proton fraction ($y_p$) substantially affects a larger volume of the stellar core, i.e. capable of sustaining rapid DU cooling. We further show that rapid rotation suppresses DU cooling by reducing the central density and $y_p$, thereby shrinking the DU-active core. This suppression is more pronounced for MPE than for DD2. Our results demonstrate that $K^-$ condensation, finite entropy, and rotation jointly exert a strong influence on the conditions for rapid neutrino cooling in NSs.
\end{abstract}

\begin{keyword}
neutron stars  \sep Bayesian analysis \sep
DDRMF \sep neutron star cooling \sep Direct Urca \sep finite temperature \sep isentropic EOS 

\end{keyword}
\end{frontmatter}

% \linenumbers

\section{Introduction}
Neutron stars (NSs) provide a unique laboratory for studying strongly interacting matter at densities far beyond those accessible in terrestrial experiments. Their cores may reach several times the nuclear saturation density ($\rho_0$), where the composition of matter remains highly uncertain. Among the possible exotic phases, antikaon ($K^-$) condensation is energetically favoured when the in-medium kaon energy ($\omega_{K^-}$) falls below the electron chemical potential \cite{Banik2001a,Banik2001b}. The condensates replace leptons to maintain charge neutrality, soften the equation of state (EoS), and reduce the star's maximum mass \cite{Pons2001,Banik2014}. The critical density for $K^-$ condensation lies in the range $2$--$5~\rho_0$, depending on the depth of attractive $U_K$ \cite{Banik2014}. Since $U_K$ remains poorly constrained \cite{Garcia-Recio, Ryu2006}, its uncertainty propagates directly to NS observables such as masses and radii.

NS cooling is a central probe of the internal composition and neutrino-emission mechanisms operating at supranuclear densities, thermal evolution observations can provide complementary constraints on the dense-matter EoS and the emergence of exotic phases \cite{Yakovlev2004}. In particular, the onset of efficient neutrino-emission channels such as the direct Urca (DU) process can dramatically accelerate cooling, making the thermal history of NSs highly sensitive to the proton fraction ($y_p$) and the presence of additional degrees of freedom in the stellar core \cite{Yakovlev2004ASR}.

The discovery of massive pulsars such as PSR J1614$-$2230 and PSR J0348$+$0432 established a lower bound of about $2\,M_\odot$ on the maximum NS mass, excluding many excessively soft EoSs \cite{J1614,j0348}. More recently, NICER observations of PSR J0030$+$0451 and PSR J0740$+$6620 have provided simultaneous mass-radius constraints \cite{Riley_j0030,Miller_j0030,Riley_j0740,Miller_j0740}, probing the density regime where exotic degrees of freedom may appear. Together, these observations may provide valuable constraints on the dense-matter EoS and the $U_K$.

Motivated by these developments, recent Bayesian studies have explored NS EoSs with $K^-$ condensation \cite{Parmar2024}. In a previous work, we combined NICER and LIGO/Virgo observations to investigate the impact of $K^-$ condensation on NS structure and $f-$mode oscillations \cite{Debanjan2025}. In the present work, we extend that analysis to examine the consequences of the inferred $K^-$ sector for the DU and NS cooling.

To explore these effects, we construct EoSs spanning the range $U_K \in [-180, -60]$ MeV, within two distinct parametrisations of a density-dependent relativistic mean-field (DDRMF) model. The onset of $K^-$ condensation increases the $y_p$ \cite{THORSSON1994693}, thereby modifying the threshold for the DU process\cite{Beznogov2023b}. The high-density behaviour of the symmetry energy further influences whether cooling proceeds through conventional nucleonic modified urca reactions or $K^-$--induced DU processes \cite{Kubis_2003,Lattimer1991}. Rapid rotation can suppress DU cooling by reducing the central density ($\rho_c$) and the corresponding central proton fraction ($y_{p,c}$) in the stellar core \cite{Negreiros2012,Beznogov2023a,Krastev2008}. Since both, $U_K$ and rotation strongly affect the $y_p$, their interplay is expected to have important consequences for NS cooling. As the NS is born in an adiabatic environment \cite{Pons2001}, an isentropic profile is chosen for studying considerably hot NS.

The paper is organised as follows. We first briefly discuss the two DDRMF parametrisations and the DU process, before presenting our results. Finally, we summarise our findings and conclude.

\section{DDRMF models}
The relativistic mean-field (RMF) model is employed to describe dense matter in the core of a NS, which consists of neutrons, protons, electrons and muons \cite{Dutra2014}. The meson-exchange framework, originally introduced by Walecka \cite{serot1986}, has since been extended to incorporate isospin asymmetry relevant for NS matter. 

The nucleons with mass $m_N$ interact via $j=\sigma$,
$\omega$, $\rho$-mesons. $g_{j N}$'s are the nucleon (N)-meson (j) couplings which are calibrated to reproduce the properties such as the charge radii, the binding energies, and the available neutron radii of several spherical nuclei at $\rho_0$ \cite{Dutra2014}. However, the core densities of a NS can reach several times this value, modeling the high-density part is still a challenge.  In DDRMF models, the in-medium modification of nucleon-meson interactions is incorporated through density-dependent couplings \cite{Hempel_2010, Malik2022}. Several prescriptions have been proposed depending on the choice of the underlying density variable, including scalar ($\rho_s$), vector($\rho$), and mixed density dependencies. This has been systematically explored for nuclear matter \cite{Typel2018} and NSs \cite{Armen&Wei2026, aprajita2025}.
  Among these prescriptions, models with scalar-density dependence for all couplings $g_{jN}$ are known to exhibit numerical and thermodynamic inconsistencies \cite{aprajita2025}. We therefore focus on the vector density-dependent DD2 and mixed density-dependent MPE parametrisations. In MPE, only the $\sigma$-meson coupling $g_{\sigma N}$ depends on the $\rho_s$, while the remaining couplings depend on the  $\rho$.

\section{Urca Process}

The DU process is one of the most efficient mechanisms for neutrino emission and rapid cooling in the core of NSs through $\beta$ decay of neutrons and electron capture on protons: $n\rightarrow p + e^- + \bar \nu_e, \quad p + e^- \rightarrow n + \nu_e$. When both processes take place, the matter attains $\beta$-equilibrium, i.e. $\mu_n = \mu_p +\mu_e$, where $\mu_i$ refers to the chemical potential of the $i$-th particle species: $i=(n,p,e)$. At temperatures below the Fermi temperature ($T_F \sim 10^{12}$ K), the participating fermions occupy states close to their respective Fermi surfaces, and their momenta can be approximated by the corresponding Fermi momenta ($k_F$). The neutrinos and antineutrinos, having negligible momenta compared to the degenerate fermions, do not significantly affect momentum balance. 
The DU process can proceed only if the triangle inequality for momentum conservation is satisfied:
 \(  k^F_p + k^F_e \geq k^F_n \) \cite{Haensel2007}, where $k^F_i$ denotes the Fermi momentum of the $i$-th particle species. The number density of each particle is given by $\rho_i= (k^F_i)^3/3 \pi^2$. This condition can be translated into a constraint on the $y_p= \frac {\rho_p}{\rho_n + \rho_p}$ which must exceed a critical value: 

 \( y_p \geq y_p^{min}\) where  \cite{Tuhin2024}
\begin{equation}
    y_{p}^{\min} = \frac{1}{1 + \left(1 + x_{e}^{1/3}\right)^{3}},
    \label{eq:DUeq}
\end{equation}
Here, \(x_e = \rho_e/(\rho_e+\rho_\mu)\), while $\rho_e$ and $\rho_\mu$ are electron and muon densities, respectively. In typical NS matter, the $y_p$ is often too low to satisfy the DU condition, thereby suppressing rapid neutrino cooling \cite{Sedrakian2024}. However, the presence of exotic degrees of freedom, such as $K^-$ condensates, hyperons, or deconfined quark matter \cite{Wang2026}, can increase the $y_p$ and enable the onset of the DU process once $y_p$ reaches approximately $11$--$15$\% \cite{Lattimer1991}. In this work, we focus on the effects of $K^-$ condensation on DU process. 

In addition to the zero-temperature $(T = 0)$, we study isentropic  matter which usually correspond to non zero temperature $(T>0)$. In our work we consider finite entropy $(S =1)$ for which the core temperature is reported around {$T$} $\approx 50$ MeV \cite{Batra2018,tuhin2021}. Throughout this work, the Boltzmann constant ($k_B$) is set to unity i.e. $k_B \equiv 1$.

\section{Results}
\subsection{Equations of state}
\noindent We choose two nucleonic ($np$) DDRMF models, DD2 and MPE, based on Refs. \cite{char2014, Armen2023} for this work. 
We also consider $K^-$ condensates ($npK^-$), which are expected to emerge at high densities and provide an alternative mechanism for ensuring charge neutrality and softening of the EoS. An additional $\phi$ meson is considered for $K^-N$ interaction. All the vector meson couplings are calculated at $\rho_0$ using the $SU(6)$ quark model as done in Refs. \cite{Batra2018, char2014}. The scalar $\sigma$ meson couplings are calculated from $U_K$ values.

We start by analysing the posterior probability distribution function (PDF) of $U_K$ for the DD2 and MPE parametrisations as shown in Fig.~\ref{pdf}. The posterior has been obtained by imposing constraints based on mass-radius inferences from NICER observation of PSR J0030$+$0451 and PSR J0740$+$6620. $U_K$ is sampled from a uniform prior spanning the range $[-180, -60]$ MeV (grey shaded region in Fig.~\ref{pdf}), as suggested by kaonic atom data \cite{Friedman1994} and theoretical estimates of in-medium $K^-$ interactions.
Both parametrisations indicate a strongly attractive in-medium interaction corresponding to $U_K = -108.94^{+32.97}_{-34.16}$ MeV for DD2 and $U_K = -119.64^{+40.96}_{-42.43}$ MeV for MPE, consistent with the onset of $K^-$ condensation inside NS interiors.
The width of the posterior distributions indicates that astrophysical observations used in this work provide only weak constraints on $U_K$.
The DD2 model exhibits a bimodal structure with peaks near $U_K \approx -130$ MeV and $-75$ MeV,  whereas the MPE parametrisation shows greater support for more negative values of $U_K$, 
indicating a model-dependent preference for deeper $U_K$.

% ~~~~~~~~~~~~~~~~~~~~~~~~~~~~~~~~~~~~~~~~~~~~~~~~~~~~~~~~~~~~~~~~~~~~~~~~~~~~~~~~~~~~~~~
\begin{figure}
    \centering
    \includegraphics[width=\linewidth]{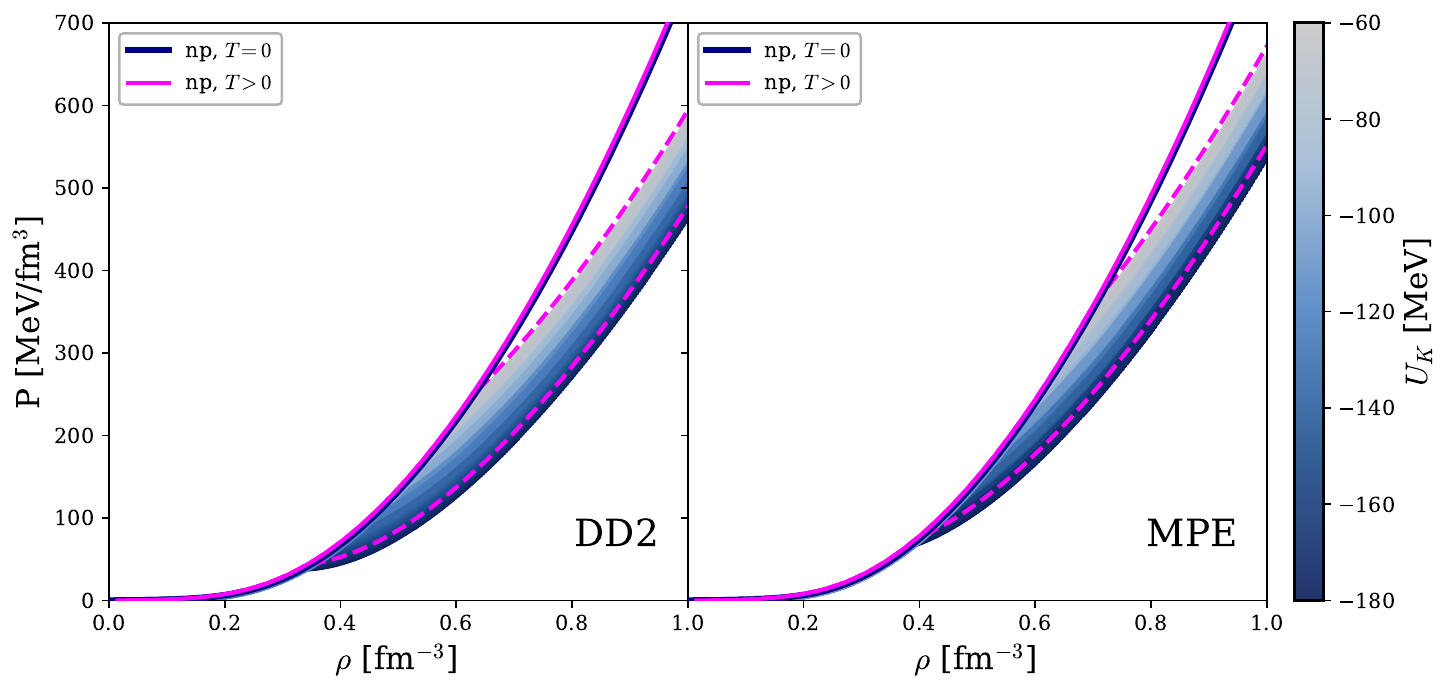}
    \caption{\label{EoS} Plot of equation of state (EoS)---pressure ($P$) for a given baryon number density ($\rho$)---for DD2 model (left panel) and MPE model (right panel) parameterisations. For cold matter ($T=0$), dark blue solid curves represent $np$ models while the colored band corresponds to the EoSs for $npK^-$ matter in the posterior. The color bar indicates the $K^-$ optical potential $U_K$. The isentropic ($T > 0$) EoSs are shown in magenta: the solid curves correspond to the $np$ case, while the dashed curves represent $npK^-$ for the extreme values of $U_K$ in the posterior.}
    \label{eos}
\end{figure}

To understand this preference, we examine the EoS directly.
Fig.~\ref{EoS} shows the pressure as a function of number density for the DD2 (left panel) and MPE (right panel) parametrisations, including the effects of $K^-$ condensation in both panels. For both cases, a unified crust is used at low densities \cite{Hempel_2010}. While different density dependencies lead to similar behavior around $\rho_0$, they produce significant variations in the EoS at supra-nuclear densities, thereby impacting NS properties. The comparative study of $np$ NS was conducted by Sedrakian et al. \cite{Armen2023}. The dark blue solid lines in Fig.~\ref{EoS} correspond to cold, $np$ matter, while the shaded bands represent the posterior distributions of the EoS in the presence of $K^-$ mesons for different values of $U_K$.
More attractive values of $U_K$ produce a larger reduction in pressure. The isentropic ($T > 0$) EoS is shown by the magenta curves. For both parametrisations, finite entropy increases the pressure relative to the cold case, although this thermal stiffening is partially offset by the softening associated with $K^-$ condensation.
As expected, more negative values of $U_K$ lead to earlier condensation onset. For a given $U_K$, the onset density ($\rho_{thres}$), listed in Tab.~\ref{kt}, in MPE is systematically higher than in DD2---this is consistent with the softer nature of the DD2 $np$ EoS. In the softer DD2 EoS, the collective behaviour of the fields leads to a faster reduction of the in-medium $\omega_{K^-}$. As a result, the threshold condition $\omega_{K^-} = \mu_e$ is satisfied at a lower density in DD2 compared to MPE, where $\mu_e$ is the chemical potential of electrons. The $\rho_{thres}$ is slightly pushed to higher density for isentropic ($T > 0$) matter, which can be attributed to the thermal effects on the chemical potentials and composition. These differences occur within the density range relevant to NS interiors and therefore influence the stellar composition and cooling behaviour.

\begin{figure}
\centering
\includegraphics[width=\linewidth]{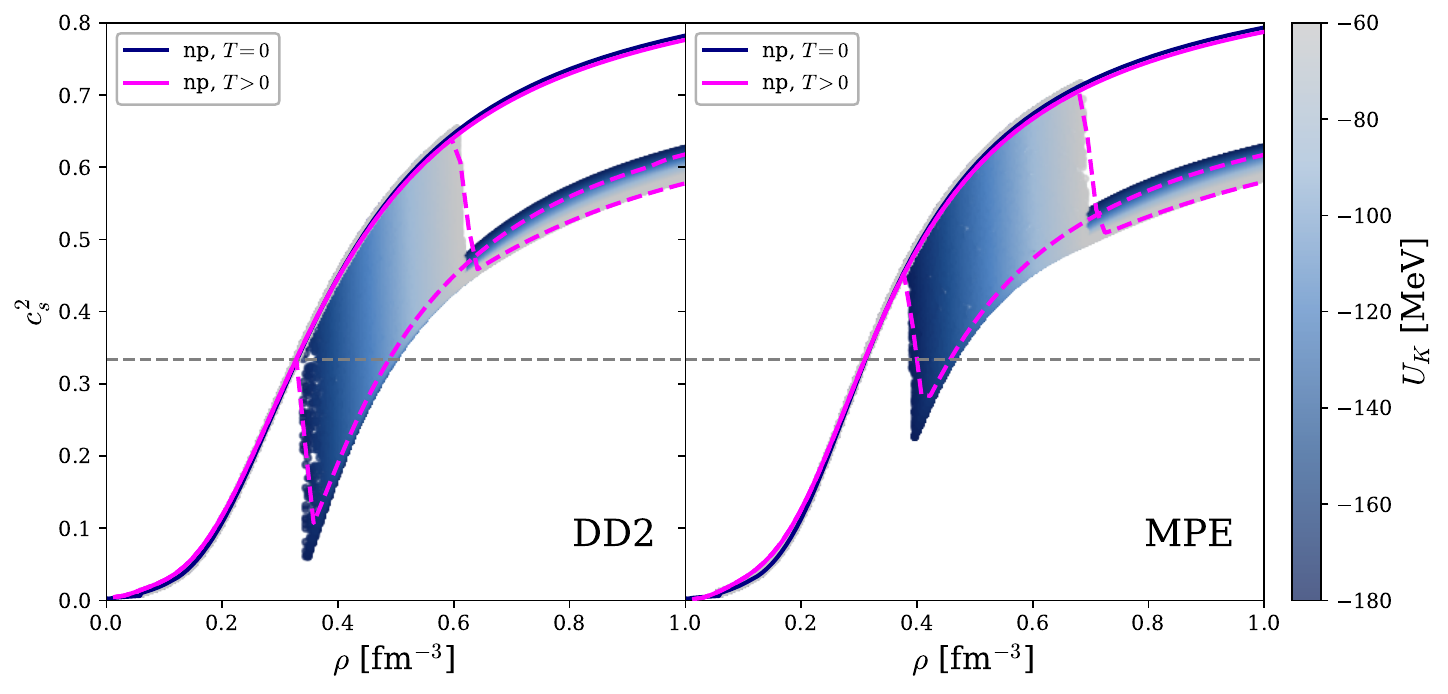}
\caption{\label{SS_SZR} Speed of sound squared ($c_s^2$), as a function of baryon number density $\rho$ for the DD2 (left panel) and MPE (right panel) parametrisations, including the effects of $K^-$ condensation. Color schemes and linestyles are same as Fig.\ref{eos}. 
The horizontal dashed line indicates the conformal limit $c_s^2 = 1/3$. All cases remain within the causal limit $c_s^2 \leq 1$.}
\label{sos}
\end{figure}
% ~~~~~~~~~~~~~~~~~~~~~~~~~~~~~~~~~~~~~~~~~~~~~~~~~~~~~~~~~~~~~~~~
\subsection{Speed of Sound}
To further assess the physical viability of the EoS, we examine the causality condition through the speed of sound, defined as $c_s^2=\partial P/\partial\epsilon$, where P and $\epsilon$ are pressure and energy density respectively. Fig.~\ref{sos} shows $c_s^2$ as a function of baryon number density for the DD2 (left panel) and MPE (right panel) parametrisations. The onset of $K^-$ condensation is marked by a dip in $c_s^2$, reflecting the associated softening of the EoS. More attractive values of $U_K$ lead to an earlier onset of condensation and a more pronounced reduction in $c_s^2$, consistent with the $\rho_{thres}$ listed in Tab.~\ref{kt}. The finite-entropy ($T > 0$) results show a systematic enhancement over the cold case due to thermal effects. At higher densities, the curves converge, indicating that the influence of $U_K$ gradually diminishes, although a residual spread persists. The horizontal dashed line denotes the conformal limit, $c_s^2=1/3$. Both parametrisations remain well within the causal limit ($c_s^2 \leq 1$), supporting their use in the subsequent analysis of NS composition and cooling.

\subsection{Proton fraction and Direct Urca threshold}
\label{subsec:proton_frac_DU_threshold}
\begin{figure}
    \centering

    \begin{subfigure}{0.49\linewidth}
        \centering
        \includegraphics[width=\linewidth]{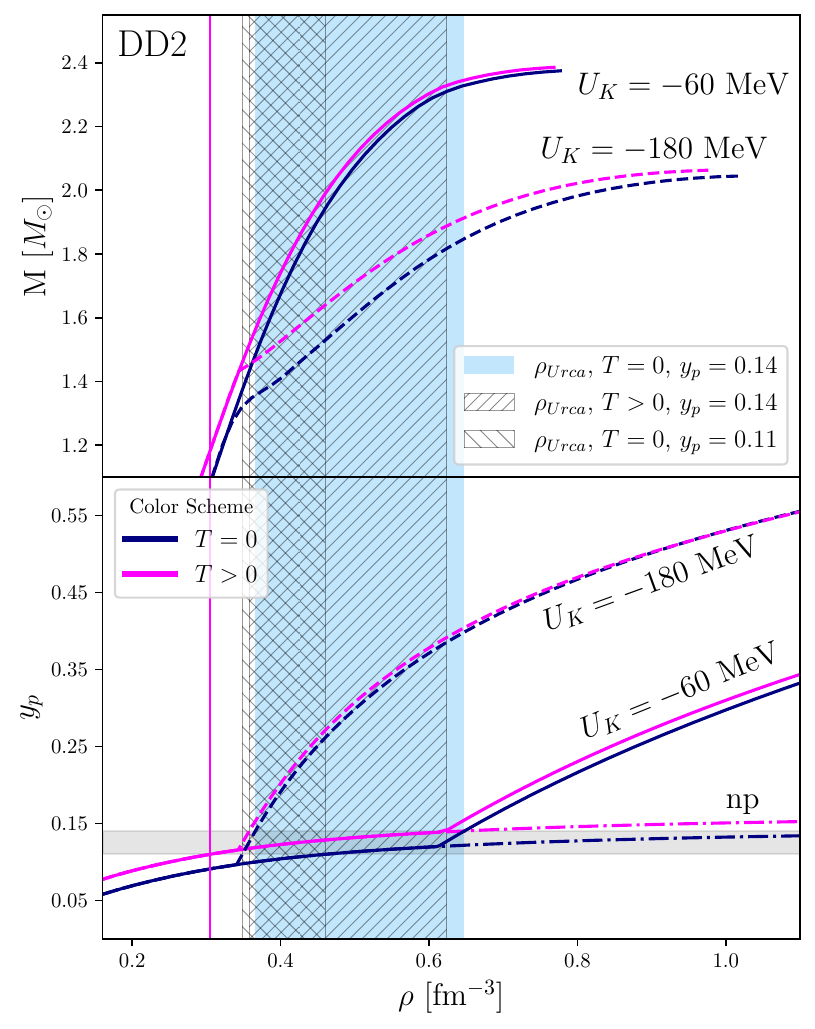}
        % \caption{}
        \label{fig:dd2_yp_rho}
    \end{subfigure}
    \hfill
    \begin{subfigure}{0.49\linewidth}
        \centering
        \includegraphics[width=\linewidth]{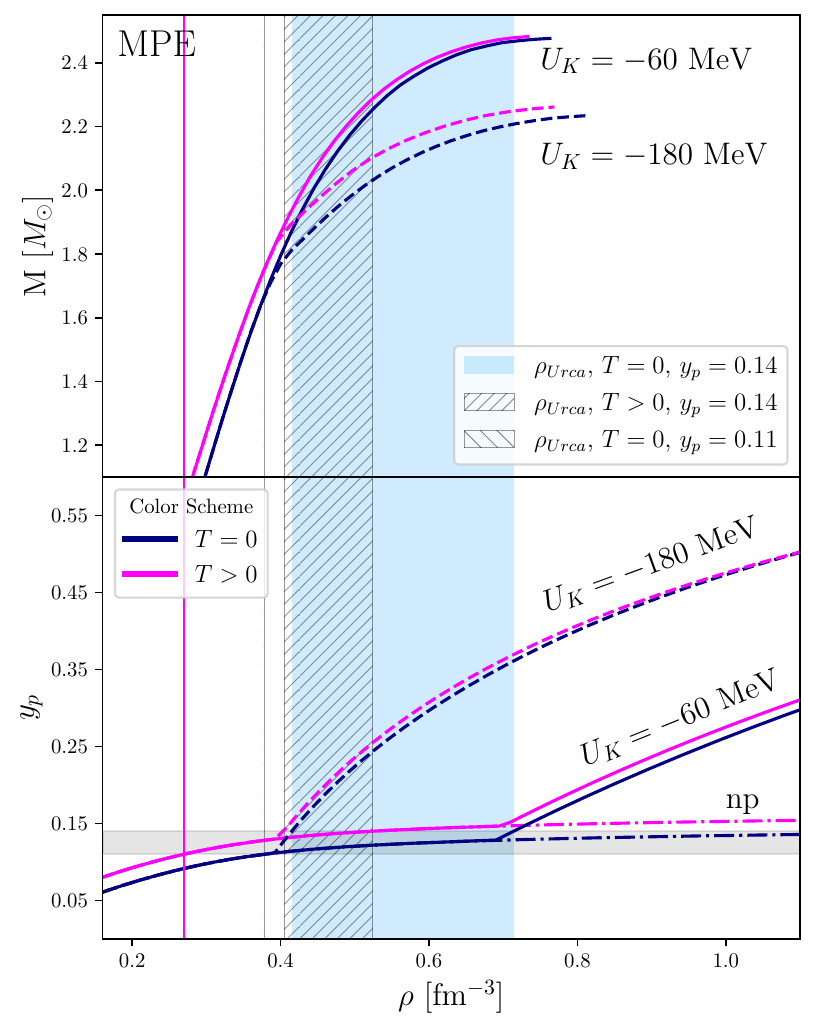}
        \label{fig:DD2_yp_vs}
    \end{subfigure}
    \caption{The bottom panels correspond to proton fraction ($y_p$) for the DD2 and MPE parametrisations, shown as functions of baryon number density ($\rho$).
    We compare the $y_p$ curves for $np$ (dash-dotted lines) and $npK^-$ matter with $U_K=-60$ MeV (solid lines) and $U_K=-180$ MeV (dashed lines).
    The dark blue (magenta) curves represent the cold, $T=0$ (finite-entropy, $T > 0$) configurations.
    The horizontal gray band denotes the threshold $y_p$ range (see beginning of Sec. \ref{subsec:proton_frac_DU_threshold} for details) for the onset of the DU cooling.
    The vertical bands (shaded and hatched) indicate the corresponding range of threshold densities for the cases considered.
    The top panels show the static NS mass configurations for $npK^-$ matter.
    The vertical bands are extended to the mass curves to determine stellar masses corresponding to the chosen DU threshold for the subsequent rotational analysis.}
    \label{fig:DD2_subplot}
\end{figure}

Having discussed the impact of $K^-$ condensation on the EoS, we now turn to its implications for NS cooling.
Fig. \ref{pf_dd2_mpe} shows the variation of $y^{min}_p / y_p$, where $y^{min}_p = 0.14$ with $\rho$ for DD2 and MPE parametrisation.
In Fig. \ref{fig:DD2_subplot}, we choose $y_p=0.11$ and $y_p=0.14$, represented by a horizontal gray band, as the threshold proton fractions above which rapid cooling through DU reactions becomes possible, based on the Ref. \cite{Lattimer1991}.
Then, for all the EoS models in this work, we examine the variation of $y_p$ with $\rho$ and find out the density $\rho_{Urca}$ at which the $y_p$ threshold for DU cooling is achieved. We compare the $\rho_{Urca}$ trends between $T=0$ and $T>0$ configurations.

% -------------------------------------------------------------

In the bottom panels of Fig. \ref{fig:DD2_subplot}, we plot the $y_p$ as a function of $\rho$ for our sets of EoSs: (i) $np$, and (ii) $npK^-$ with $U_K=-60$ MeV and $U_K=-180$ MeV. The cold ($T=0$) cases are represented by dark blue colors, with magenta curves representing the finite-entropy ($T > 0$) configurations. The $y_p$ is systematically higher for all finite-entropy cases, compared to that in $T=0$ case.

\begin{table*}
    \centering
    \begin{tabular}{ccccccccc}
    \hline\hline
         Model & $y_p$ & $U_K$ (MeV) & $M_{Urca}$ ($M_\odot$) & $M_{grav}$ ($M_\odot$) & $\rho_{Urca}$ (fm$^{-3}$) & $\rho_{thres}$ (fm$^{-3}$)& $\nu_{Urca}$ (Hz)\\
         \hline
         DD2 & $0.14$ & $-60$ & $2.329$ ($2.239$) & $2.359$ ($2.359$) & $0.647$ ($0.624$) & $0.614$ ($0.620$) & $418.87$ ($432.26$) & \\
         DD2 & $0.14$ & $-180$ & $1.356$ ($1.453$) & $1.386$ ($1.483$) & $0.366$ ($0.358$) & $0.345$ ($0.343$) & $340.90$ $(333.50)$ &\\
         \hline
         DD2 & $0.11$ & $-60$ & $1.942$ ($1.183$) & $1.972$ ($1.213$) & $0.460$ ($0.305$) &  & $364.89$ $(325.04)$ \\
         DD2 & $0.11$ & $-180$ & $1.320$ ($1.183$) & $1.350$ ($1.213$) & $0.348$ ($0.305$) & & $330.97$ $(325.04)$ \\
         \hline
         MPE & $0.14$ & $-60$ & $2.468$ ($2.287$) & $2.478$ ($2.297$) & $0.714$ ($0.524$) & 0.690 (0.700) & $225.59$ ($207.25$)\\
         MPE & $0.14$ & $-180$ & $1.812$ ($1.870$) & $1.822$ ($1.880$) & $0.415$ ($0.405$) & 0.393 (0.396)& $205.99$ $ (222.57)$\\
         \hline
         MPE & $0.11$ & $-60$ & $1.666$ ($1.011$) & $1.676$ ($1.021$) & $0.378$ ($0.269$)  &  & $180.51$ ($190.12$)  \\
         MPE & $0.11$ & $-180$ & $1.664$ ($1.011$) & $1.674$ ($1.021$) & $0.378$ ($0.269$) &  & $190.90$ ($190.12$) \\
         \hline\hline
    \end{tabular}
    \caption{$M_{Urca}$ is the NS mass corresponding to the baryon number density ($\rho_{Urca}$) at which the DU process becomes feasible. A column for the $K^-$ onset density ($\rho_{thres}$) is included for comparison. $M_{grav}$ is the mass of NS chosen for calculating the variation of the central baryon density with rotation. The rotational frequency at which DU cooling capability is lost is denoted by $\nu_{Urca}$ and reported in Hz. All entries in the table are for $T =0$ $(T>0)$.}
    \label{krastev_table}
\end{table*}

For $np$ matter (dash-dotted lines), the DU threshold at $y_p = 0.14$ is not attained in the cold ($T=0$) configuration for either parametrisation, consistent with the general expectation for cold $np$ matter~\cite{Tuhin_2021_apj}.
In contrast, the threshold is reached for the $T > 0$ configurations---at $\rho = 0.641$ fm$^{-3}$ and $0.524$ fm$^{-3}$ for DD2 and MPE models respectively.
At the lower threshold $y_p = 0.11$, however, both models satisfy the DU condition even in cold matter--- at $\rho = 0.460$ fm$^{-3}$ and $0.378$ fm$^{-3}$ for DD2 and MPE, respectively.
Once again, $T > 0$ configurations reach the DU threshold $y_p = 0.11$ easily.
Interestingly, the corresponding $\rho_{Urca}$ values---$0.305$ fm$^{-3}$ (DD2 case) and $0.269$ fm$^{-3}$ (MPE case) are lower than the lowest $K^-$ onset densities (corresponding to $U_K = -180$ MeV)---$0.343$ fm$^{-3}$ (DD2 case) and $0.396$ fm$^{-3}$ (MPE case).
This implies that appearance of $K^-$ for $T > 0$ configuration is not necessary to achieve the $y_p = 0.11$ threshold.

For cold ($T= 0$) matter, the inclusion of $K^-$ condensates significantly alters the situation at $y_p = 0.14$.
More attractive values of $U_K$ casue earlier threshold crossings.
The corresponding threshold densities $\rho_{Urca}$ are listed in Tab.~\ref{krastev_table}.
The blue shaded band denotes the $T=0$ threshold range for $y_p = 0.14$, spanning $0.366$--$0.647$ {fm}$^{-3}$ for DD2 and $0.415$--$0.714$ fm$^{-3}$ for MPE.
The slightly broader band for the MPE case indicates a stronger sensitivity of the DU onset to $U_K$.
Finite entropy lowers the threshold further, as indicated by the inclined hatched (forward slash) region, covering densities between $0.358$--$0.624$ fm$^{-3}$ for DD2 and $0.405$--$0.524$ fm$^{-3}$ for MPE.
The results differ considerably for $y_p = 0.11$. For cold matter, the $npK^-$ EoSs within the DD2 parametrisation predict $\rho_{Urca}$ between $0.348$--$0.460$ fm$^{-3}$ corresponding to $U_K$ between $-180$ MeV and $-60$ MeV.
This is depicted by the hatched (backward slash) band. This spread vanishes completely for MPE parametrisation---$\rho_{Urca} = 0.378$ fm$^{-3}$ is independent of the $U_K$. However, $\rho_{Urca}$ is independent of $U_K$ for both models ($1.213$ fm$^{-3}$ and $1.021$ fm$^{-3}$ for DD2 and MPE cases, respectively) corresponding to the $T  > 0, y_p = 0.11$ case. 

The top panels in Fig. \ref{fig:DD2_subplot} show the NS mass as a function of $\rho$ at its center for the same set of EoSs as in the lower panels. 
The stellar masses ($M_{Urca}$), for static configurations, corresponding to these DU thresholds are obtained by noting the intersection among the mass curves and the corresponding vertical lines extended from the bottom panels.
These values are mentioned in Tab. \ref{krastev_table} in the column for $M_{Urca}$.
Densities of matter inside the NSs with mass less than $M_{Urca}$ for a given EoS will never reach $\rho_{Urca}$. As a result, such star will never undergo DU cooling.
Finite entropy supports larger maximum masses owing to the marginally stiffer thermal EoS, while softer EoS with $U_K=-180$ MeV corresponds to a smaller maximum mass star.
The trends for $M_{Urca}$ mimics the trend in $\rho_{Urca}$---spread in $M_{Urca}$ exists when a spread in $\rho_{Urca}$ exists for a given $y_p$ and entropy, and $M_{Urca}$ becomes independent of $U_K$ when $\rho_{Urca}$ is constant.
The differences between DD2 and MPE arise from their distinct density dependence of the nucleon-meson couplings---$g_{jN}$---leading to different $y_p$ evolution at suprasaturation densities.

\begin{figure}[htbp]
    \centering
    \begin{subfigure}{0.49\linewidth}
        \centering
        \includegraphics[width=\linewidth]{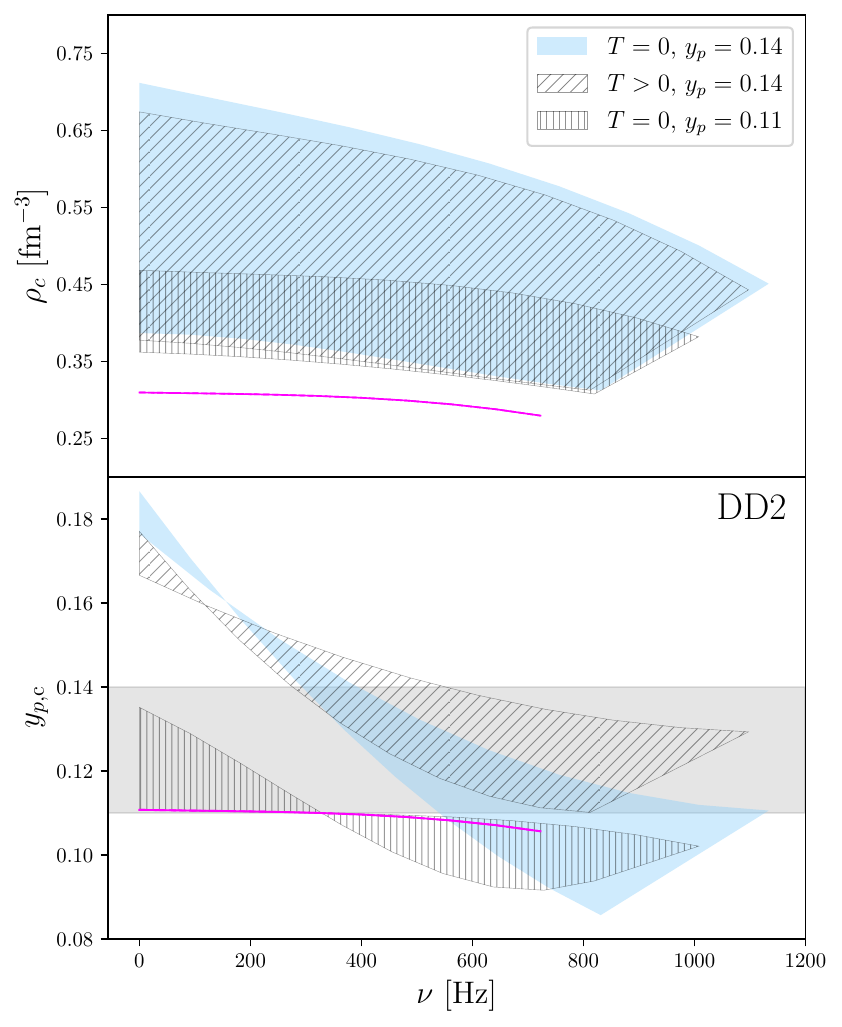}       
    \end{subfigure}
    \hfill
    \begin{subfigure}{0.49\linewidth}
        \centering
        \includegraphics[width=\linewidth]{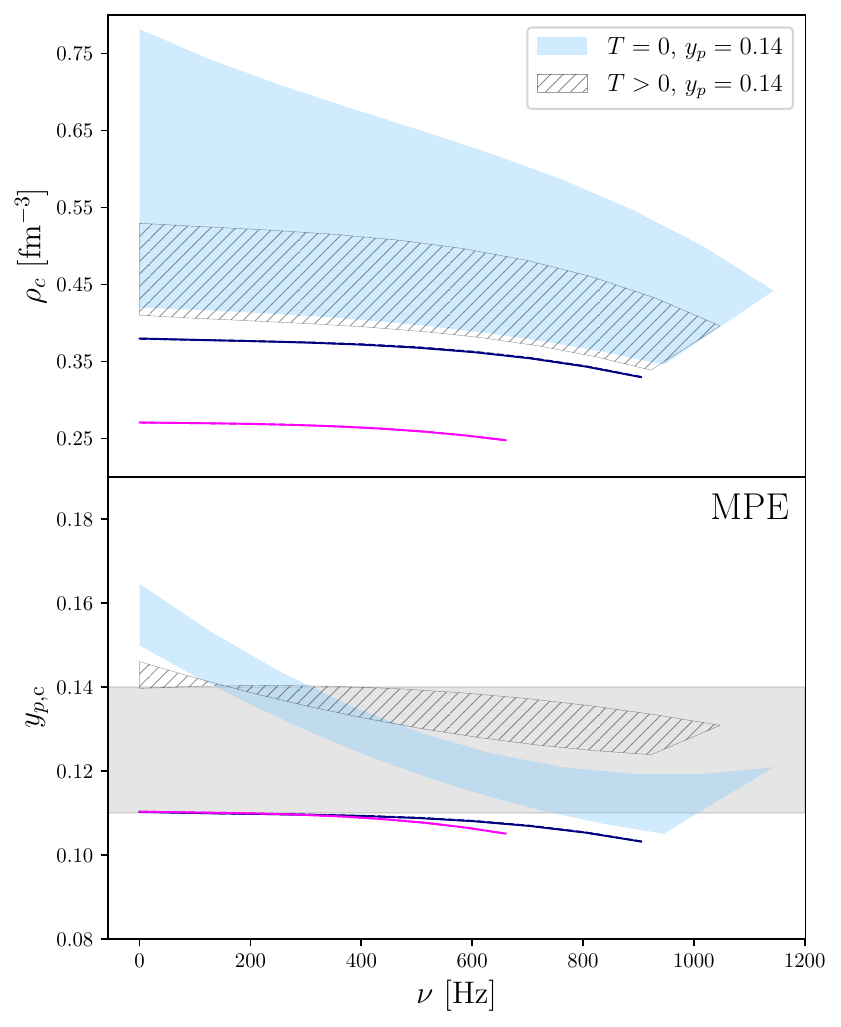}     
    \end{subfigure}
    \caption{Central baryon density ($\rho_c$) (top) and central proton fraction ($y_{p,c}$)(bottom) as functions of rotational frequency $\nu$ (in Hz) for the DD2 (left) and MPE (right) parametrisations.
    In all the panels, the bands span a region corresponding to data for $U_K \in [-180, -60]$ MeV--- 
    blue solid band: $T=0$, $y_p = 0.14$, hatched (forward slash) band: $T > 0$, $y_p = 0.14$, and hatched (vertical) band: $T=0$, $y_p = 0.11$.
    For DD2, the magenta curve in top and bottom panels correspond to $T>0$, $y_p = 0.11$ sequences with the $U_K = -60$ MeV and $U_K = -180$ MeV cases overlapping.
    The horizontal gray band marks the DU threshold range. Each sequence runs from the static limit to the Keplerian limit ($\nu_K$) of the frequency regime.
    }
    \label{fig:rotation}
\end{figure}
    
\subsection{Rapid rotation and Direct Urca cooling}
Having discussed the static configurations and the onset of the DU process, we now examine the effects of rapid rotation on the composition and cooling properties of NSs within the DD2 and MPE parametrisations.
Rotating stellar sequences are constructed using the {\tt RNS} code for stationary and axisymmetric compact stars in general relativity.
Fig.~\ref{fig:rotation} shows the $\rho_c$ (top panels) and central proton fraction ($y_{p,c}$) (bottom panels) as functions of rotational frequency for fixed gravitational-mass ($M_{grav}$) sequences, selected to lie near and above the $M_{Urca}$ identified in the previous section, are listed in Tab.~\ref{krastev_table}. Each sequence extends from the static case ($\nu=0$) to the Keplerian frequency limit $\nu_K$, beyond which mass shedding occurs.

As the rotational frequency increases, centrifugal support reduces $\rho_c$, and consequently $y_{p,c}$ for both the models. Stars that satisfy the DU condition in the static limit may therefore fall below the threshold at sufficiently high rotation rates, suppressing rapid neutrino cooling.
These rotation rates are identified as $\nu_{Urca}$ (refer to the last column in Tab.~\ref{krastev_table} for values).
This effect is strongest for configurations lying close to the DU threshold at $\nu=0$.

The shaded regions represent the range of stellar sequences obtained by varying $U_K$ from $-180$ to $-60$ MeV. The blue solid band corresponds to ($T=0$, $y_p = 0.14$), the hatched (forward slash) band corresponds to ($T > 0$, $y_p = 0.14$), and the hatched (vertical) band to ($T=0$, $y_p = 0.11$). The magenta curves denote the ($T > 0$, $y_p = 0.11$) sequences with $M_{grav}=1.213~M_\odot$ (DD2) and $1.021~M_\odot$ (MPE).

For DD2, the blue band ($T=0$, $y_p = 0.14$) starts well above the DU threshold and its lower boundary crosses below the threshold at high rotational frequencies, indicating that some configurations lose their DU cooling capability as the spin increases. The hatched (forward slash) band ($T=0$, $y_p = 0.11$) lies closer to the threshold and falls below it at comparatively lower frequencies. For MPE, the hatched (vertical) band is absent because the $y_p = 0.11$ threshold is already reached in $np$ matter before the onset of $K^-$ condensation. The blue band in MPE case starts closer to the DU threshold than in DD2, making the MPE sequences more susceptible to rotational suppression of DU cooling. The magenta curves, corresponding to the lowest-mass finite-entropy configurations, are the most sensitive to rotation and fall below the threshold even at moderate spin frequencies  (refer to the last column in Tab.~\ref{krastev_table}).

A comparison of the two parametrisations reveals a clear EoS dependence. The bands are systematically broader for DD2, reflecting its larger spread in threshold densities across the allowed $U_K$ range. In contrast, the finite-entropy MPE sequences exhibit a steeper decline in $y_p$ with increasing rotation, indicating a stronger centrifugal suppression of DU cooling.

DD2 generally supports higher rotation rates and exhibits a broader variation with entropy and $U_K$ than MPE.
For both parametrisations, the $T > 0$ sequences reach slightly larger frequencies before their $y_{p,c}$ goes below the threshold than their $T = 0$ counterparts. The lower-mass $y_p = 0.11$ sequences cross the threshold at lower frequencies.
In MPE, the $y_p = 0.11$ sequences show only weak sensitivity to either $U_K$ or entropy, indicating that the $\nu_{Urca}$ for these stars is largely independent of the $K^-$ optical potential.
Overall, rapid rotation significantly modifies the core composition and cooling behaviour of $K^-$--condensed NSs. The extent of DU suppression and the $\nu_{Urca}$ spin rate depend sensitively on the underlying EoS, thermodynamic conditions, and $U_K$, with DD2 generally supporting a broader range of compositions and higher rotation rates than MPE.

% @@@@@@@@@@@@@@@@@@@@@@@@@@@@@@@@@@@@@@@@@@@@@@@@@@@@@@@@@@@
\begin{figure}[htbp]
    \centering
    \begin{subfigure}{0.495\linewidth}
        \centering
    \includegraphics[width=\linewidth]{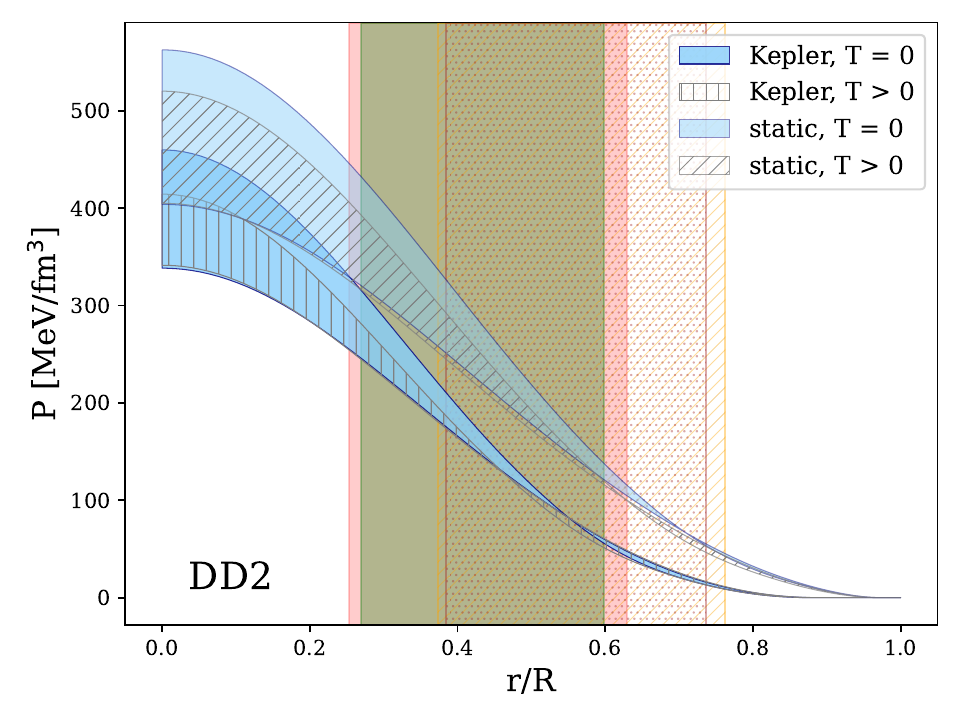}
    % \caption{}
        \label{fig:DD2_pressure_profile}
    \end{subfigure}%
    % \hfill
    \hspace{-1.9mm}
      \begin{subfigure}{0.495\linewidth}
        \centering
        \includegraphics[width=\linewidth]{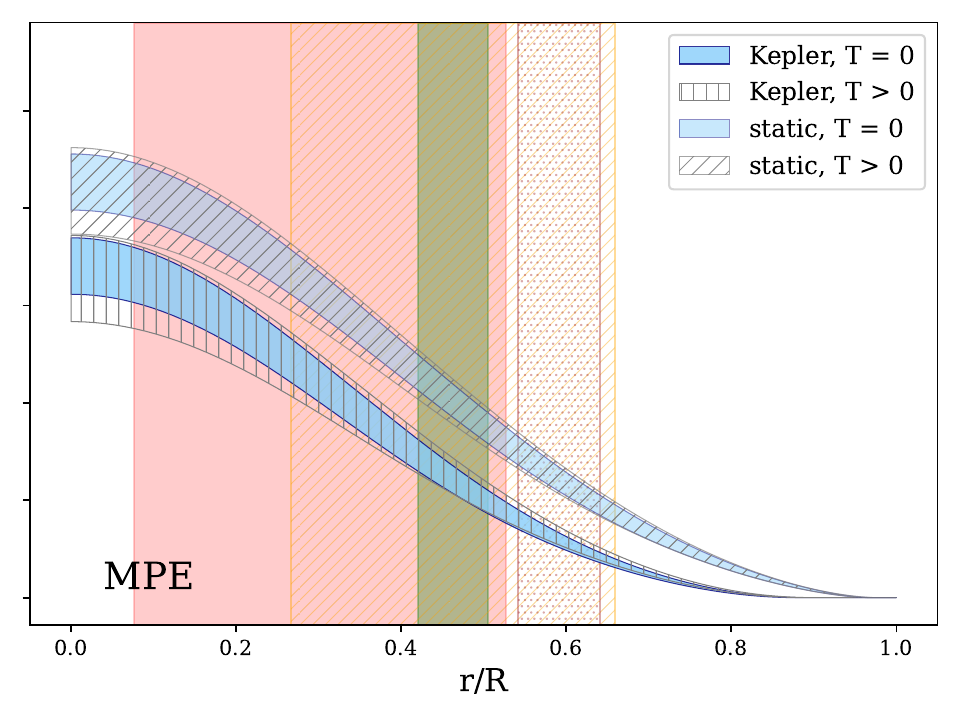}
        % \caption{}
        \label{fig:MPE_pressure_profile}
    \end{subfigure}
    \caption{Pressure profiles as a function of the normalized radial coordinate $r/R$ ($R$ is the stellar radius) for the maximum-mass configurations of DD2 (left panel) and MPE (right panel). We compare static and Keplerian sequences for $T = 0$ and $T>0$ cases.
    The vertical bands mark the range of DU onset locations corresponding to $U_K \in [-180, -60]$ MeV, with the left and right boundaries of each band corresponding to $U_K = -60$ MeV and $U_K = -180$ MeV, respectively.
    The vertical hatched bands correspond to the static configurations---yellow forward slash hatches ($T = 0$), brown dotted hatches ($T > 0$). The vertical solid bands---peach ($T = 0$), green ($T > 0 $) correspond to the Keplerian configurations for the radius at $\rho_{Urca}$.
    The spread of each band reflects the sensitivity of the DU onset to the $U_K$.}    
\label{fig:pressure_profile}
\end{figure}

Now, we now examine the spatial extent of the stellar region in which rapid neutrino cooling via the DU process can operate. 
In the absence of the DU process, NS cooling is known to be governed by the less efficient Modified Urca \cite{Haensel1995}. The enhancement of the $y_p$ by $K^-$ condensation, together with thermal effects in finite-entropy matter, shifts the DU threshold to lower densities and enlarges the region of the stellar core where rapid neutrino emission can operate. 

Fig.~\ref{fig:pressure_profile} shows the pressure profile as a function of normalized radial coordinate $r/R$ ($R$ stands for stellar radius), for the maximum-mass configurations of DD2 (left panel) and MPE (right panel), comparing static and Keplerian sequences.
Here, colored vertical bands represent the radial extent of the DU region at different $U_K$ for static and Kepler configuration.
More attractive values of $U_K$ shift the DU threshold to larger radial fractions, thereby increasing the volume of the core available for rapid cooling. The limiting radius decreases for the Keplerian configuration in both $T = 0$ \& $T > 0$ cases, which shows that rapid rotation reduces the DU-active volume.
This behaviour is consistent with the reduction of $\rho_c$ caused by centrifugal support, which on the other hand lowers $y_p$ and makes the DU condition hard to satisfy as we have seen earlier in Fig.~\ref{fig:rotation}.  In contrast, rotation pushes the onset inward, reducing the DU-active region as a consequence of the lower $\rho_c$ in rapidly rotating stars. This rotational suppression is more pronounced for MPE than for DD2, consistent with the stronger response of the stiffer MPE EoS to centrifugal deformation near the Keplerian limit.

\subsection{Summary \& Discussions}

In this work, we have investigated the onset of antikaon ($K^-$) condensation in neutron star (NS) cores 
using two density-dependent relativistic mean-field (DDRMF) parametrisations---DD2 and MPE---which differ in their density dependence of nucleon-meson couplings. The $K^-$ optical potential $U_K$, which governs the threshold density for condensation, was treated as a free parameter over the range $[-180, -60]$ MeV and constrained via Bayesian inference using NICER mass-radius observations of PSR J0030$+$0451 and PSR J0740$+$6620.

The posterior distributions of $U_K$ favour strongly attractive in-medium $K^-$ interactions for both parametrisations. However, the distributions remain broad, reflecting the limited sensitivity of the astrophysical constraints in this work to $U_K$. While DD2 exhibits a slightly bimodal posterior structure, MPE shows a preference for comparatively deeper (more attractive) potentials. Future high-precision radius measurements and gravitational-wave observations are expected to provide significantly tighter constraints on the $K^-$ sector.

The equation of state (EoS) analysis shows that more attractive values of $U_K$ lead to an earlier onset of 
condensation, a larger reduction in pressure, and a softer EoS. The onset density is systematically lower in DD2 than in MPE, reflecting differences in the underlying density dependence of the nucleon-meson couplings. Finite entropy ($T > 0$) slightly pushes the onset to higher densities due to thermal modifications of the chemical potentials. All configurations remain within the causal limit $c_s^2 \leq 1$ and show a characteristic dip in the sound speed at the condensation onset, supporting the physical viability of the EoS across the full parameter range.

A central finding of this work is the strong sensitivity of the Direct Urca (DU) threshold to both $U_K$ and the thermodynamic conditions. The $K^-$--induced enhancement of the $y_p$ shifts the DU onset to lower densities, with the threshold density ranging from $0.366$--$0.647$ fm$^{-3}$ for DD2 and $0.415$--$0.714$ fm$^{-3}$ for MPE at $T=0$. 
Finite entropy lowers these thresholds further. An interesting exception is found for MPE at $y_p = 0.11$ where the DU condition is already satisfied in cold nucleonic matter, rendering the threshold largely insensitive to $U_K$. These results highlight the importance of both the $K^-$ sector and thermal effects in determining the cooling behaviour of NS.

We further demonstrate that rapid rotation can qualitatively modify the cooling behaviour by suppressing DU emission in stars that satisfy the threshold in the static limit. As the spin frequency increases toward the Keplerian limit, centrifugal support lowers the central density and $y_p$, potentially suppressing DU cooling in stars that would otherwise cool rapidly. This rotational suppression is more pronounced for MPE than DD2, and is strongest for low-mass finite-entropy configurations lying close to the DU threshold at zero rotation. The pressure profile analysis confirms that more attractive $U_K$ values enlarge the DU-active core volume, while rotation systematically shrinks it. The interplay between $K^-$ condensation, entropy, and rotation thus introduces a rich parameter-dependence in the cooling history of NSs.

While several aspects of this work merit further investigation, including extensions to other density-dependent parametrisations and the incorporation of additional exotic degrees of freedom, the present analysis already demonstrates that the interplay of $K^-$ condensation, finite entropy, and rapid rotation can substantially alter the onset and spatial extent of rapid neutrino cooling, highlighting the importance of these effects in determining the cooling behaviour of NSs.

\section*{Acknowledgments}
AS and SB acknowledge the Department of Science \& Technology, Govt. of India, for the support via project no: CRG/2022/008360. AS gratefully acknowledges Anagh Venneti, Skund Tewari for valuable assistance and insightful suggestions that contributed to this work.

\bibliographystyle{elsarticle-num}
\bibliography{biblio}

\clearpage
\onecolumn
\appendix
\section{Supplementary Material}

\begin{figure}[htbp]
    \centering
    \includegraphics[width=0.60\linewidth]{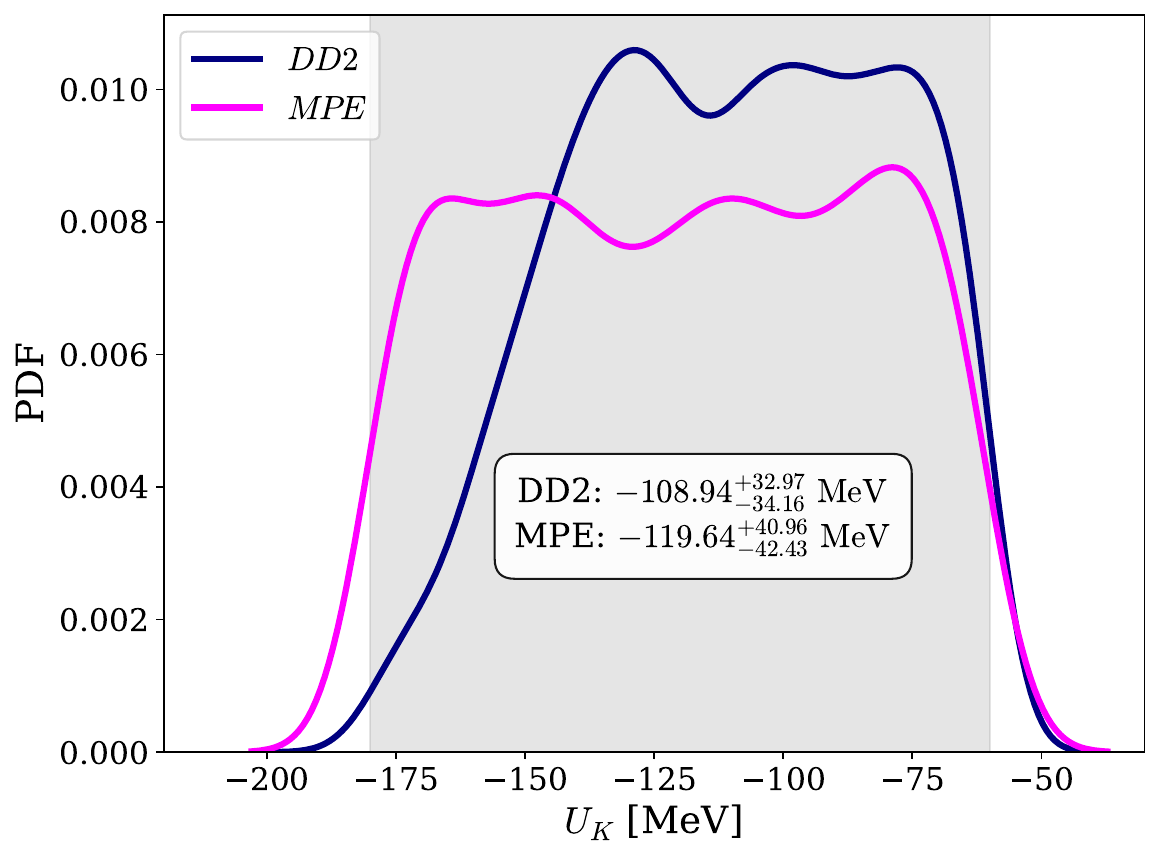}
    \caption{\label{PDF}Posterior probability density distributions of the antikaon optical potential 
    $U_K$ for different DDRMF models showing the effect of constraints from NICER data for PSR J0030+0451 and PSR J0740+6620. The grey-shaded region is the uniform prior corresponding to the range $[-180,-60]$ MeV. The curves have been obtained using kernel density estimates for the DD2 (dark-blue) and MPE (magenta) parametrisations.}
    \label{pdf}
\end{figure}
% ~~~~~~~~~~~~~~~~~~~~~~~~~~~~~~~~~~~~~~~~~~~~~~~~~~~~~~~~~~~
\begin{table}[htbp]
\centering
\caption{$K^-$ condensate threshold density (in fm$^{-3}$) for different optical potentials $U_K$ (MeV). }
\label{kt}
\setlength{\tabcolsep}{4pt}
\begin{tabular}{lccccccc}
    \hline\hline
    Model & \multicolumn{7}{c}{$U_K$ (MeV)} \\
    \cline{2-8}
          & $-180$ & $-160$ & $-140$ & $-120$ & $-100$ & $-80$ & $-60$ \\
    \hline
    DD2 (T = 0) & 0.345 & 0.377 & 0.417 & 0.461 & 0.510 & 0.562 & 0.614 \\
    MPE (T = 0) & 0.393 & 0.437 & 0.481 & 0.530 & 0.582 & 0.634 & 0.690 \\
    DD2 (T > 0) & 0.343 & 0.379 & 0.420 & 0.464 & 0.513 & 0.565 & 0.620 \\
    MPE (T > 0) & 0.396 & 0.439 & 0.485 & 0.535 & 0.587 & 0.642 & 0.700 \\
    \hline\hline
\end{tabular}
\end{table}
%~~~~~~~~~~~~~~~~~~~~~~~~~~~~~~~~~~~~~~~~~~~~~~~~~~~~~~~~~~~~~

\begin{figure}
    \centering
    \includegraphics[width=\linewidth]{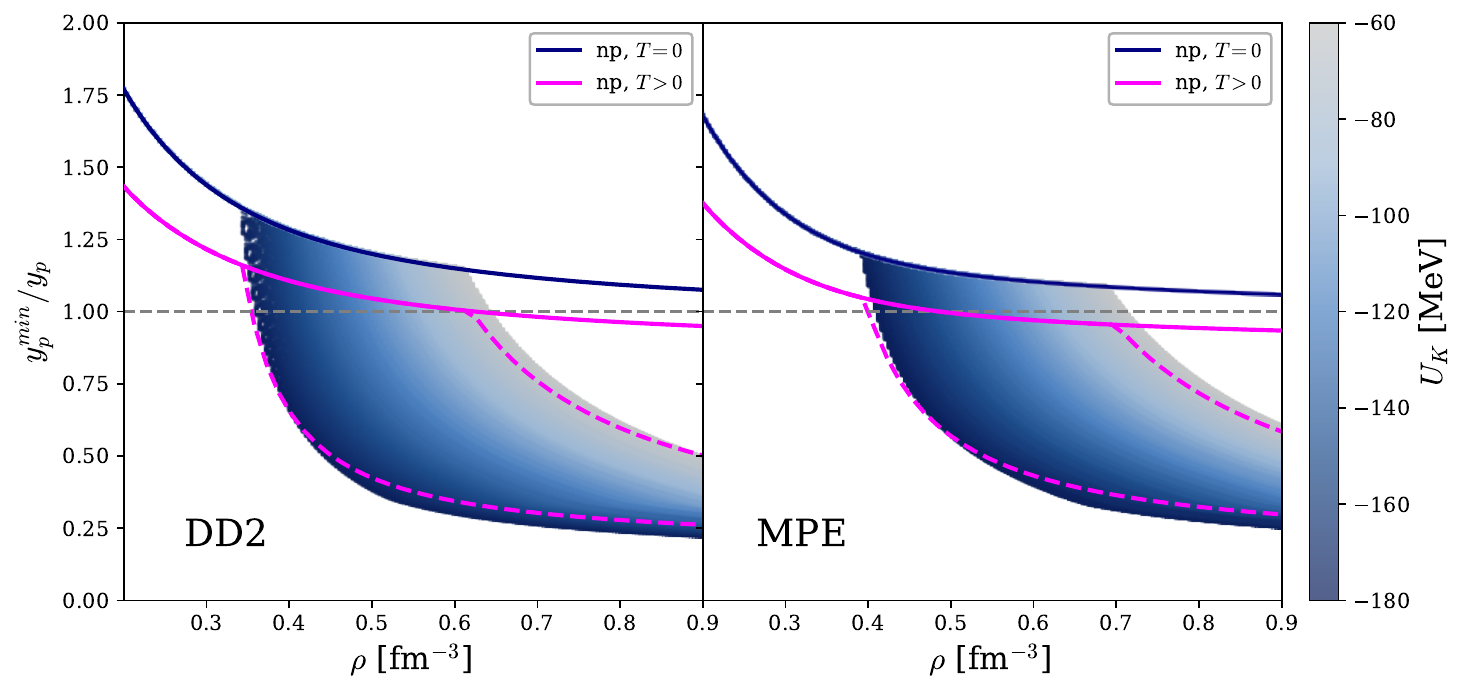}
    \caption{Density dependence of $y_p^{\min}/y_p$ for DD2 (left panel) and MPE (right panel) parameterisations for different $K^-$ optical potential $U_K$. Here $y_p$ refers to the proton fraction at a given baryon number density ($\rho$).}
    \label{pf_dd2_mpe}
\end{figure}

\end{document}